# From visibility to vulnerability: how women scientists face gendered hostility in science communication

**Maider Eizmendi-Iraola[1] and Simón Peña-Fernández[2]***

1. University of the Basque Country (UPV/EHU) (ROR: 000xsnr85)
   maider.eizmendi@ehu.eus / https://orcid.org/0000-0002-5894-8238
2. University of the Basque Country (UPV/EHU) (ROR: 000xsnr85)
   simon.pena@ehu.eus / https://orcid.org/0000-0003-2080-3241

* Corresponding author

**Abstract**: Public communication of science has become a central component in the relationship between science and society. However, the media exposure of research staff has given rise to new forms of hostility, especially in digital environments. The study Experiences of researchers who interact with the media and social networks in Spain [Science Media Centre Spain, 2024] investigates — via a survey (N=237) — the incidence and typology of these attacks in the Spanish context. More than half of the research staff (51.05%) reported experiencing negative incidents, with a higher prevalence among women (56.9%) than men (46.2%). The attacks differ by gender: women face more challenges regarding their scientific capacity and sexist remarks, whereas men are more frequently targeted over their professional integrity. These dynamics reveal structural gender biases that affect the wellbeing and legitimacy of female scientists, emphasising the need for institutional policies with a gender perspective.

## 1. Introduction

Public science communication has moved from being a peripheral activity within academia to becoming a foundational pillar in the relationship between science and society. In a context characterised by global challenges, communicating science clearly, rigorously and accessibly has become indispensable for fostering an informed citizenry and evidence-based decision-making [Bucchi & Trench, 2021].

The use of digital media and social networks has broadened the interaction between scientists and the public, facilitating knowledge exchange and the development of digital communities that promote scientific literacy and critical thinking [Martin Neira et al., 2023]. Theoretically, social media platforms can also enhance the visibility of women scientists by helping them overcome the obstacles they have traditionally faced when seeking access to mainstream media as expert sources [Sugimoto & Larivière, 2023; Loverock & Hart, 2018; Mena-Young, 2018]. Nonetheless, they have also exposed scientists to new forms of public scrutiny and aggression [O'Grady, 2022; Nogrady, 2021; Nölleke et al., 2023; Global Witness, 2023].

For this reason, attacks on scientists engaged in public outreach have become a recurring theme in contemporary studies of science communication and science-society relations — particularly in the wake of the COVID-19 crisis, which brought this phenomenon into sharper focus, revealing its significant personal and social repercussions. [O'Grady, 2022; Nogrady, 2021].

In this vein, in the Spanish context, the study Experiences of researchers who interact with the media and social networks in Spain [Science Media Centre Spain, 2024] constitutes one of the first empirical approaches to the impact of media exposure on the country's scientific community. From a methodological standpoint, a quantitative approach was adopted, using a self administered survey distributed via email. This approach removes the need for an interviewer, enabling faster data collection and increasing confidence in anonymity, which may encourage more open responses.

The study population comprises all experts in science and technology areas who have been contacted by Science Media Centre (SMC) España as sources of information from the start of its operations in 2022 until July 2024. This ensures that all participants have engaged at least once in this type of activity — though with varying degrees of intensity (N = 1,405). A total of 237 individuals completed the survey. The response rate was 17%, which is high compared with earlier studies on this issue. Moreover, the abandonment rate — one of the main problems posed by online surveys — was only 5.5 %.

The typology of attacks analysed in this study is based on predefined closed-response categories included in the questionnaire. These categories were established prior to data collection and were informed by previous research on hostility and harassment in digital environments. Therefore, no post hoc qualitative coding procedure was applied to classify incidents. The full questionnaire can be consulted at the following address: https://zenodo.org/records/18639293.

The main objective of this study is to analyse researchers' experiences after communicating about science in the media, with particular attention to the incidence and typology of negative incidents, as well as their consequences and the measures adopted in response to them. All this is approached from a gender perspective and focused on the Spanish context. This is therefore an exploratory study that has sought to bring to the fore a reality which must necessarily be studied more deeply given the personal and social implications, and to extend it to other territories — provided that the specific environment under analysis is taken into account.

## 2. Incidence and typology of the attacks

Overall, the data indicate that the scientific community holds a predominantly positive (58.65%) or very positive (24.47%) perception of their participation in the media and particularly value the opportunity to convey their message to the public and to increase the visibility of their research.

However, the data also show that more than half of the researchers surveyed (51.05%) have experienced one or more forms of attack after engaging in science communication over the past five years. Overall, the incidence of attacks on scientific personnel is substantial, although the prevalence is noticeably higher in the case of women: 56.9% of women researchers reported having encountered some form of hostility after communicating the results of their work, compared with 46.2% of men.

This gap confirms that media exposure carries a differential risk for women, who face more intense scrutiny. It should be emphasised here that this higher frequency of attacks against women in digital environments is not limited to the field of science. In Spain, women who express themselves publicly — such as professionals in politics or communication — receive more attacks on social media [Piñeiro-Otero & Martínez-Rolán, 2021] and, in general, this hostility tends to be more intense when they hold opinions or participate in fields historically considered masculine [Zamora-Martínez et al., 2024].

In this broader context of escalating hostility in the public sphere, these data must be situated within a context characterised by what has been described as a "war on science" [Hardy et al., 2019], as well as by attacks on professionals working in other fields, such as journalism [Kim & Shin, 2025; Peña-Fernández et al., 2025] or politics [Martín-Gutiérrez et al., 2025].

Beyond frequency, this study shows how gender shapes the nature and consequences of such attacks. Building on previous studies that have analysed hostility and harassment against scientists in digital environments [Amarasekara & Grant, 2019; Nogrady, 2021; O'Grady, 2022; Global Witness, 2023], and drawing on information gathered by SMC staff through their interactions with scientists, a range of different types of attacks has been analysed, including: (1) insults; (2) comments about physical appearance; (3) remarks concerning origin, ethnicity, ideology, religion, or beliefs; (4) comments about sexual orientation or gender identity; (5) remarks questioning professional competence; (6) comments undermining professional integrity; (7) persistent and repetitive contact; (8) publication of personal information; (9) threats of physical or sexual violence; and (10) death threats.

The findings suggest that men tend to report criticism related to their credibility or potential conflicts of interest — a way of calling their scientific integrity into question — whereas women report a significantly higher proportion of attacks targeting their scientific competence (34.31% of women compared to 24.24% of men), primarily through the social network X and via comments on media outlets and online platforms.

This difference is meaningful: it points to a symbolic structure that continues to position scientific authority within a masculinised framework. These differences, although apparently subtle, reveal two opposing modes of delegitimising scientific authority: in men, through moral suspicion; in women, through epistemic doubt. Both mechanisms contribute to eroding public trust in scientific personnel, but they do so asymmetrically and therefore with different social effects, as will be discussed further below.

Thus, the male scientist embodies the model of the legitimised expert, whereas the female scientist appears as a figure whose authority is perceived as fragile, incomplete or conditional. Therefore, when a female scientist participates in a media debate, her testimony is evaluated through a cultural filter that associates rationality and authority with masculine attributes. Hence, this asymmetry does not operate merely as a form of personal discrimination, but as a structural mechanism that limits the recognition of knowledge produced by women.

These differences must be understood in a culture that has historically identified rationality, objectivity and authority — traits associated with science — with the masculine [Fricker, 2007;

Harding, 1986]. These stereotypes have been reflected in, and reinforced by, media content, which tends to associate scientific performance with traits traditionally assigned to men. Leadership and analytical capability are associated with masculine roles [Carli et al., 2016], qualities more valued than those attributed to women, which are more closely linked with the emotional dimension [Husu & Tainio, 2016]. Studies carried out in this field have also highlighted the presence of stereotypes associated with image [Mitchell & McKinnon, 2019; Attenborough, 2011; Chimba & Kitzinger, 2010] and with the idea of exceptionality in science [Eizmendi-Iraola & Peña-Fernández, 2023].

Furthermore, the data collected in the SMC study also show a significant difference in comments received by male and female scientists on their physical appearance, with a much higher incidence for women (8.82% of women versus 0.76% of men). This is not an isolated datum: several studies have found that female scientists receive more sexist remarks than their male peers when communicating via digital platforms [Amarasekara & Grant, 2019; Döring & Mohseni, 2019], which may discourage them from participating in such outreach activities [Cambronero-Saiz et al., 2023].

Regarding the remaining types of attacks, no statistically significant gender differences are observed overall. However, some patterns emerge: women report fewer insults than men, but more comments related to their origin, ethnicity, ideology, religion or beliefs, as well as to their sexual orientation or gender identity. They also report fewer instances of persistent and repetitive contact. In contrast, women refer slightly more often to the publication of personal data and to threats of physical or sexual violence, including death threats, although the overall prevalence of these experiences remains low.

With regard to the topics addressed by those who report having experienced negative incidents, COVID-19, climate change, vaccines and science policy stand out as areas with the highest proportion of attacks. However, these are also the most frequently discussed topics overall. Even so, a notable incidence is observed among scientists who communicate on issues related to gender and gender diversity, as well as animal welfare.

## 3. The consequences of the attacks

Gendered patterns in these attacks are also reflected in their consequences. Although the majority of attacks occur in digital environments, the impact transcends the virtual space and affects psychological wellbeing and the sense of belonging and legitimacy within the scientific community.

Indeed, the study data reveal that almost one in four respondents report that the attacks and comments they receive after disseminating their work do not affect them; however, the percentage of scientists who suffer anxiety or other psychological problems (22.18 %) or insecurity is noteworthy (19.37 %) and differentiated by gender, since in both cases women are affected in a higher proportion. In the case of personal insecurity, for example, the percentage difference is more than 15 points (27.87% versus 12.10 %). Comments questioning credibility, which affect men to a greater extent, are associated with a decline in productivity and, in some cases, with the abandonment of outreach activities. This suggests that moral suspicion may have tangible professional consequences, as it can undermine researchers’ confidence and willingness to engage publicly.

Negative experiences appear to be shaped by gender stereotypes associated with women, but also amplify them, reinforcing social norms that sanction a female voice in the public sphere and undermining their self-perception as a legitimate subject of science. In the case of men, according to the data obtained, they are more affected in terms of loss of productivity and perception of loss of professional reputation.

Since social networks are the main channels through which these types of attacks are received, the priority measures adopted in response relate to reporting and blocking users on networks (41.90 %) or ceasing to read the comments received (32.04 %). However, gender disaggregated data show that women tend to seek help in a greater proportion than men in their professional environment or institution. Thus, the close supportive networks in which women perceive proximity and trust are the spaces most appreciated by female scientists.

In this respect, in terms of the support received from their managers and employers, there is a clear gender difference: women feel supported to a greater extent than men and feel that the help has been useful (23.77% versus 7.64 %).

At this point it is important to underline, however, that a high percentage of scientists opt not to report what happened to anyone (54.92% of women and 43.31% of men) or perhaps cannot find adequate channels to do so. Hence the problem of attacks on scientists involved in dissemination is a hidden reality, despite the growing interest it generates.

## 4. Towards gender-aware science communication

In the SMC study, in addition to the questionnaire items, participants were invited to expand their responses and were asked about the measures that should be taken to prevent or respond to these attacks. In summary, scientists proposed that academic institutions should value outreach work and also adopt protective measures, taking into account the contribution they make to the public image of the institution.

> [S1] "Currently universities and public centres encourage their researchers to do science communication, but from my experience, if there is some form of harassment or insults, suddenly it's not their problem any more and the researcher is left to deal with it alone. So it seems institutions are there for the good side of outreach (the publicity) but not for the bad." (Man)
> [S2] "There is a contradiction: we are asked to be in the media, but our interventions are not valued and when we receive attacks the institution washes its hands." (Man)
> [S3] "Institutional backing from your centre is very important so that you are not delegitimised." (Woman)

Thus, given the growing importance of dissemination in academic careers and its value for the public profile of research institutions, accompanying it is vital, as is ensuring protocols and reporting channels that act swiftly.

Beyond these more general measures, the study data reveal that gender-based hostility exists in science communication. Recognising this reality is the first step, but it remains necessary to transform this finding into concrete institutional and cultural actions that allow epistemic equality and the safety of those who communicate science.

In this respect, different lines of action might be considered, such as incorporating gender analysis into the evaluation of science communication and into studies of attacks faced by the scientific community. It would thus be necessary to collect gender-disaggregated data in all communication activities or to include analytical categories that help identify differences in exposure, public response and emotional consequences produced.

With regard to the measures that institutions should adopt, it is necessary to provide tools that equip both male and female scientists to manage the emotional strain generated by different types of comments, and to offer legal support mechanisms in the most serious cases, such as death threats or threats of violence. In addition, the different activities planned to train staff in science communication should include content related to gender inequality in science, offering tools to identify and respond to comments or attacks based on gender stereotypes. In addition, creating support networks in which female scientists share their experiences and develop joint strategies to create safe environments for their outreach activity is beneficial — taking into account that, as the study reveals, proximal environments are the preferred channels for female scientists. These networks can mark the difference between a hostile experience that isolates and one that empowers.

## 5. Discussion

Hostility towards science is neither a marginal nor a transient phenomenon: it constitutes one of the major challenges of contemporary science communication. However, its impact is not neutral. The data from the report Researchers' experiences in their interaction with the media and social networks show that the experience of hostility is deeply mediated by gender. Incidence is greater among women, and qualitatively it is also uneven: women are more often questioned about their scientific capacity; men, about their honesty. Therefore, knowledge produced by women continues to need proving, while that of men remains presumed.

It is also crucial to consider the broader social consequences of such attacks. Hostility and emotional strain inevitably shape which topics are researched, who dares to speak publicly, and what kind of science reaches the public sphere. This hostility and harassment, therefore, do not only affect individual science communicators, but also the scientific communication ecosystem itself, diminishing its diversity — particularly in relation to socially contested or sensitive topics.

The study also brings to light a contradiction within the current academic culture: while it promotes a model of science communication based on self-promotion and competition for public attention, it fails to recognise the emotional toll and social risks involved. Institutions demand media visibility from their researchers, yet fall short when it comes to offering effective protection against such attacks.

Recognising these dynamics requires critically rethinking gender stereotypes associated with women, particularly in science. Moving towards a more inclusive scientific culture involves not only training scientists in communication skills, but also educating society to listen to diverse voices with equal credibility. Equity in science communication will not be achieved simply by adding more women to the conversation, but by transforming the conditions that determine who is granted legitimacy and who is able to speak without fear of attack. As such, any measures adopted to protect the scientific community when engaging in public communication must take these asymmetries into account and seek to ensure that science communication genuinely becomes a means of enhancing the visibility of women scientists, reshaping the norms of recognition and credibility in the public sphere.

Despite the findings of this study, it is important to acknowledge its limitations. The sample consists of individuals who have participated in outreach activities through the Science Media Centre, a group that is already active and committed to science communication. It would therefore be appropriate to examine the experiences reported by scientists who engage through other dissemination channels. In addition, analysing a larger sample would provide greater robustness and consistency to the findings.

## Acknowledgments

The data analyzed in this article were originally collected as part of a technical report commissioned by the Spanish Foundation for Science and Technology (FECYT) [Science Media Centre Spain, 2024].